\documentclass[%
 reprint,
 amsmath,amssymb,
 aps,
]{revtex4-2}

\usepackage{graphicx}
\usepackage{dcolumn}
\usepackage{bm}
\usepackage{ dsfont }
\usepackage{ bbold }
\usepackage{comment}

\usepackage{tikz-feynman}
\tikzfeynmanset{compat=1.1.0}
\usepackage{hyperref}
\usepackage[capitalize]{cleveref}

\newcommand\cmt[1]{}
\newcommand{\leri}[1]{\left(#1 \right)}
\newcommand{\sinc}[1]{\text{sinc}\left(#1 \right)}

\begin{document}

\preprint{APS/123-QED}

\title{Nelson-Barr ultralight dark matter}

\author{Michael Dine}
\affiliation{
    Santa Cruz Institute for Particle Physics and Department of Physics,\\University of California, Santa Cruz,\\
    Santa Cruz, CA, USA
}

\author{Gilad Perez}
\affiliation{Department of Particle Physics and Astrophysics,\\
Weizmann Institute of Science, Rehovot, Israel 7610001}

\author{Wolfram Ratzinger}
\affiliation{Department of Particle Physics and Astrophysics,\\
Weizmann Institute of Science, Rehovot, Israel 7610001}

\author{Inbar Savoray}
\affiliation{Berkeley Center for Theoretical Physics, University of California, Berkeley, CA 94720, USA}
\affiliation{Physics Division, Lawrence Berkeley National Laboratory, Berkeley, CA 94720, USA}

\begin{abstract}
We show that, in the Nelson-Barr solution to the strong CP-problem, a naturally light scalar can arise. It gives rise to a completely new phenomenology beyond that of the celebrated QCD axion, if this field constitutes dark matter, as the CKM elements vary periodically in time.
We also discuss how the model can be tested using quantum sensors, in particular using nuclear clocks, which leads to an interesting synergy between different frontiers of physics. 
\end{abstract}

\maketitle

%%%%%%%%%%%%%%%%%%%%%%%%%%%%%%%%%%%%%%%%%%%%%%%%%%%%%%%%%%
\section{\label{sec:introduction}Introduction}
%%%%%%%%%%%%%%%%%%%%%%%%%%%%%%%%%%%%%%%%%%%%%%%%%%%%%%%%%%
There are three widely discussed solutions to the strong CP problem.  These are the possibility of a massless $u$ quark, for some time ruled out by lattice gauge theory, the Peccei-Quinn solution~\cite{Peccei:1977hh}, and the possibility of spontaneous breaking of CP, which we will refer to as the Nelson-Barr solution~\cite{Nelson:1983zb,Nelson:1984hg,Barr:1984qx} (for earlier attempts see~\cite{Mohapatra:1978fy,Georgi:1978xz}).  In this work, we focus on aspects of the latter.  In particular, the minimal realization of the Nelson-Barr solution involves an additional vector-like fermion, as well as a single complex scalar that spontaneously breaks the CP symmetry~\cite{Bento:1991ez}. 
Some special structure is required to obtain a working model, which can be the result of a discrete or (non-anomalous) approximate continuous symmetry. As we show below, the symmetry could render the phase of the scalar field comparatively light since it is only broken by the standard model (SM) quark flavor-mixing. In this paper, we explore the possibility that this boson is {\it extremely} light and can be a viable ultralight dark matter (UDM) candidate.
The same symmetry
suppresses otherwise dangerous contributions to the strong CP phase. Unlike the case of the QCD axion the light field has linear scalar couplings to quarks, which arise as a result of the fact that the CKM angles and phase depend linearly on the UDM field.
The requirement that it does not lead to unobserved long range forces places a lower bound on the scale of
symmetry breaking, but at the same time opens up the possibility to an observable time-dependent signal in flavor factories or future atomic and nuclear clock experiments. 
%\vspace*{-.2cm}

%%%%%%%%%%%%%%%%%%%%%%%%%%%%%%%%%%%%%%%%%%%%%%%%%%%%%%%%%%
\section{\label{sec:model}A minimal Nelson Barr model}
%%%%%%%%%%%%%%%%%%%%%%%%%%%%%%%%%%%%%%%%%%%%%%%%%%%%%%%%%%
In this paper we study the minimal model of Ref.~\cite{Bento:1991ez}. It introduces an additional vector-like quark pair $q$ ($\bar q$) that carries the same (opposite) SM charge as the right-handed up-quark, in addition to a neutral complex scalar, $\Phi=(f+\rho)\exp(i\theta)/\sqrt2$\,. The Lagrangian contains the following couplings
\begin{equation}
    {\cal L} \supset \mu \bar q q + (g_{i} \Phi+\tilde g_{i}\Phi^*) \bar u_{i} q+ y^{u}_{ij} \tilde{H}Q_i \bar u_{j}+ y^{d}_{ij} H Q_i \bar d_{j}+ \dots \label{eq:lagrangian}
\end{equation}
It is further assumed that the theory is CP conserving, and CP is only spontaneously broken by the  expectation value of $\Phi$\,, such that $\mu,y^u$ and $y^d$ are real. %The CP violating expectation value of $\Phi=f/\sqrt{2}\exp(i\theta)$ is generated by new physics. 
In the presence of the Higgs vacuum expectation value, $v$\,, the up-quark mass is given by the following $4\times 4$ matrix:
\begin{equation}
    {\cal M}_u = \left ( \begin{matrix} \mu & B  \cr 0 & m_u \end{matrix} \right ),\ \ m_u = y^uv,~B_i=(g_{i} \Phi+\tilde g_{i}\Phi^*)\,.
    \label{eq: 4times4 quark matrix}
\end{equation}
Due to the absence of the bottom left entry, and the fact that $\Phi$ only appears in the off-diagonal entry of the above matrix, $\arg(\det({\cal M}_u))=0$ holds and no QCD phase is introduced (while the CKM phase is unconstrained).  This can be ensured by introducing an additional $\mathds{Z}_2$ symmetry under which $q,~\overline{q}$ and $\Phi$ are odd while the SM fields are even. An approximate U(1) flavor symmetry can also enforce this structure instead, which also protects the mass of the phase of $\Phi$ (see more details below).

To identify the quark masses and mixings, we focus on the structure of  ${\cal M}_u{\cal M}_u^\dagger$\,.
Assuming that the vector-like quark is heavy $\mu,|B|\gg m_u$\,, we can integrate it out and are left with an effective up-quark mass matrix $\tilde{m}_u$ satisfying
\begin{equation}
    (\tilde m_u \tilde m_u^\dagger )_{ij}=\left (  (m_u m_u^T )_{ij} - \frac{(m_u)_{ik} B^\dagger_k B_\ell (m_u^T)_{\ell j}} {\mu^2 + B_fB_f^\dagger} \right ) \,.\label{eq: effective quark mass matrix}
\end{equation}
The CKM matrix is the product of the $SU(3)$ matrix required to diagonalize $\tilde m_u \tilde m_u^\dagger$ and the left $SO(3)$ rotation required to diagonalize $y^{d}$\,. Assuming $\langle\theta\rangle={\cal O}(1)$\,, $\mu\lesssim |B|$\,, and the vectors $g$ and $\tilde g$ are of comparable magnitude and not parallel in flavor space (see discussion in~\cite{Davidi:2017gir}), the resulting CKM matrix has an $\mathcal{O}(1)$ CP violating phase.

In the rest of the paper we investigate the possibility that the potential generating the VEV of $\theta$ is shallow (again here we refer to~\cite{Davidi:2017gir}, which effectively already used this property within the relaxion framework). This corresponds to the presence of a global symmetry holding to a very good
approximation.  In this case, one finds at low energies a remaining light scalar field. The couplings of this light scalar $\phi$ are found by replacing the VEV $\theta$ with $\theta_0+\phi/f$\,. We discuss them in detail in the next sections focusing on the following set of couplings. We take $g\propto(1,0,0),\ \tilde{g}\propto (0,1,0)$\,, with $m^u$ diagonal and $m^d=V^{d\dagger}\mathrm{diag}(m_d,m_s,m_b)$\,, where $V^{d}\in SO(3)$ is the real valued CKM matrix of the original Lagrangian. We will see below that this choice gives the best prospects for detecting the model through flavor physics.

These parameters can arise naturally, for example if a global shift symmetry in $\theta$ rotates $\Phi \rightarrow e^{i \theta} \Phi$\,, and $\bar u_1 \rightarrow e^{-i \theta} \bar u_1, ~\bar u_2 \rightarrow e^{i \theta} \bar u_2$\,. This symmetry is only broken by the off-diagonal entries of the Yukawa matrices turning the angular variable $\theta$ into a pseudo-Nambu-Goldstone boson. In particular we show in \cref{sec:challenges} that loop corrections to the mass are suppressed by the Yukawas being small as well as the CKM matrix being mostly diagonal. A more detailed discussion of possible construction of the required structure can be found in \cref{sec:model building}.  The symmetries we consider are free of anomalies, so differ from a Peccei-Quinn symmetry.  
They further account for the lightness of $\phi$ as well as suppressing potentially dangerous contributions to the strong CP phase discussed in~\cite{Dine:2015jga}.

%%%%%%%%%%%%%%%%%%%%%%%%%%%%%%%%%%%%%%%%%%%%%%%%%%%%%%%%%%%%%%%%%%%
\section{\label{sec:osc_CKM}Oscillating flavour observables}
%%%%%%%%%%%%%%%%%%%%%%%%%%%%%%%%%%%%%%%%%%%%%%%%%%%%%%%%%%%%%%%%%%%
With the parameters given above one can then readily check that $\tilde m_u \tilde m_u^\dagger$ of \cref{eq: effective quark mass matrix} is diagonalized by a rotation of the first two types of up-quarks $O_{12}$ and a phase transformation $P$ removing the phase induced by the complex scalar $\Phi$
\begin{equation}
P=\mathrm{diag}\left(1,\exp\left[-2i\left(\theta_0+\frac{\phi}{f}\right)\right],1\right)\,.
\end{equation}
The resulting effective CKM matrix is given by
\begin{equation}
    V= O_{12}PV^{d}\,.\label{eq:CKM matrix}
\end{equation}
The CKM phase is directly related to the phase $\theta$ of the complex scalar, which consists of a background value $\theta_0$ and a light scalar $\phi$. If $\phi$ is light enough and constitutes dark matter, the CKM phase oscillates in time. The amplitude and period of the oscillation depend on the local dark matter density, $\rho_\mathrm{DM}\approx2.5\times10^{-6}~\mathrm{eV}^4$\,, and on the mass of the scalar, $m_\phi$:
\begin{equation}
    \frac{\phi(t)}{f}=\frac{\sqrt{2\rho_\mathrm{DM}}}{f m_\phi}\cos(m_\phi t)\,.\label{eq:uldm_osc}
\end{equation}
 Note that if $O_{12}= \mathbb{1}$ the phase can be removed. In fact the Jarlskog invariant, which all CP violating observables are proportional to, vanishes if the rotation angle $\theta_{12}$ goes to zero
\begin{align}
    J &=  \text{Im}\leri{V_{ud}V^*_{ub}V_{tb}V^*_{td}}\nonumber\\
    &= \frac{1}{2}|V_{tb}||V_{td}| \sin{2(\theta_0+\phi/f)}\sin{2{\theta}_{12}}\leri{V^{d}_{cd}V^{d}_{ub}-V_{cb}^{d}V^{d}_{ud}}\,.
\end{align}
The time independent part given by $\theta_0$  should match the observed value of $J\approx 3\times 10^{-5}$\,.
Assuming $|V|\approx |V^{d}|$\,, this leads to $\kappa\equiv\sin{2\theta_0}\sin{2{\theta}_{12}}\approx 0.2$\,. To linear order, the $\phi$ dependence of CP violating observables is therefore given by
\begin{equation}
    J = J^{0}\left(1+\frac{2}{\tan 2\theta_0}\frac{\phi}{f}\right)\,,
\end{equation}
where $J^0$ is the time independent part obtained by setting $\theta=\theta_0$\,.
In our following sensitivity estimates we assume besides $|V|\approx |V^{d}|$ that $\theta_0$ is a random phase such that $\sin{2\theta_0}\sim\cos{2\theta_0}\sim1$\,. In \cref{sec:experimental implications} we justify these simplifying assumptions.

It turns out that with our choice of parameters also the absolute value of the CKM entries involving the up and charm quark vary in time.
\begin{equation}
|V_{u/c\,j}|^2\approx|V^{\mathrm{0}}_{u/c\,j}|^2\left(1\mp 2\frac{V^{d}_{c/u\, j}}{V^{d}_{u/c\, j}}\kappa \frac{\phi}{f}\right)\,,\label{eq: CKM matrix dependence phi}
\end{equation}
where $|V^{\mathrm{0}}_{ij}|$ is  the time independent part of the CKM matrix, with $i=u,c,t\,,$ and $j=d,s,b$\,.
Note that, with our particular choice of $g$ and $\tilde g$\,, the CKM elements involving the top quark are unaffected, and are thus real and have absolute value fixed to $|V_{tj}|=|V^{d}_{tj}|$\,. This choice leads to the weakest bounds from other searches, as we show in the next section. 

We conclude that the parameters most susceptible to the oscillations of $\phi$ are $|V_{us}|$\,, $|V_{ub}|$\,, $|V_{cd}|$ and the CKM phase. For these the relative change due to the oscillation is approximately given by the amplitude of $\phi/f$ in Eq.~\eqref{eq:uldm_osc}\,. Any observable $O$ proportional to these parameters, e.g. a meson decay width, therefore also oscillates with the same relative amplitude.
The amplitude of the oscillation in these observables $\Delta O$ relative to the background value $\overline{O}$ is given as
\begin{align}
    \frac{\Delta O}{\overline{O}}&=\mathcal{O}(1)\frac{\sqrt{2\rho_\mathrm{DM}}}{f m_\phi} \nonumber\\
    &\sim10^{-5}\times\frac{10^{14}~\mathrm{GeV}}{f}\times\frac{10^{-21}~\mathrm{eV}}{m_{\phi}}\,,\label{eq: amplitude observables}
\end{align}
where we assumed that the light scalar is all of dark matter. The value of $f\sim10^{14}$ GeV saturates the bounds discussed in the next section, while $m_\phi\sim10^{-21}$ eV is at the lower limit set by cosmological and astrophysical observations (see Ref.~\cite{Hui:2021tkt} for a recent review). 

To get a sense of the achievable experimental sensitivities, we would like to estimate the error that can be achieved in a given experiment. While most experiments currently report their results under the assumption of a time-independent observable, if the time-stamped data is made available it could be possible to search for time variations through a Fourier-like analysis. However, statistical fluctuations will hinder the experimental ability to observe oscillations at a certain frequency. The spectral power of such statistical noise is frequency-independent (for the non-zero frequencies). For an experiment of length $T$ and time resolution $\Delta t$, one can show that if the phase of the oscillations is unknown, the relative amplitude of oscillations with period $T>\tau>2\Delta t$ can be constrained to the level of $\sim2/\sqrt{N_0}$, where $N_0$ is the expected total number of events in the time-independent case. For a more detailed discussion of how to perform a measurement of this kind see Refs.~\cite{Dev:2020kgz,Losada:2021bxx,Losada:2023zap}. 

As a conservative benchmark one may quote an existing study aimed at the determination of the observable, which was calculated under the assumption that it is constant in time. Such studies however impose stringent cuts on a given dataset in order to keep systematic errors under control. The number of used events is then much smaller than the total number of recorded events. Assuming that systematic errors don't vary over time there is however no need for such cuts when searching for oscillations. As an optimistic benchmark we may therefore assume that a relative statistic error $\approx2/\sqrt{N_{0}}$ can be achieved, where $N_{0}$ is the number of events an experiment has collected. Here we also note that while in many flavor observables theoretical uncertainties (such as on the hadronic parameters) are important, they are irrelevant for the uncertainty on the oscillations at a non-zero frequency.

\begin{figure}
    \centering
    \includegraphics[width=.47\textwidth]{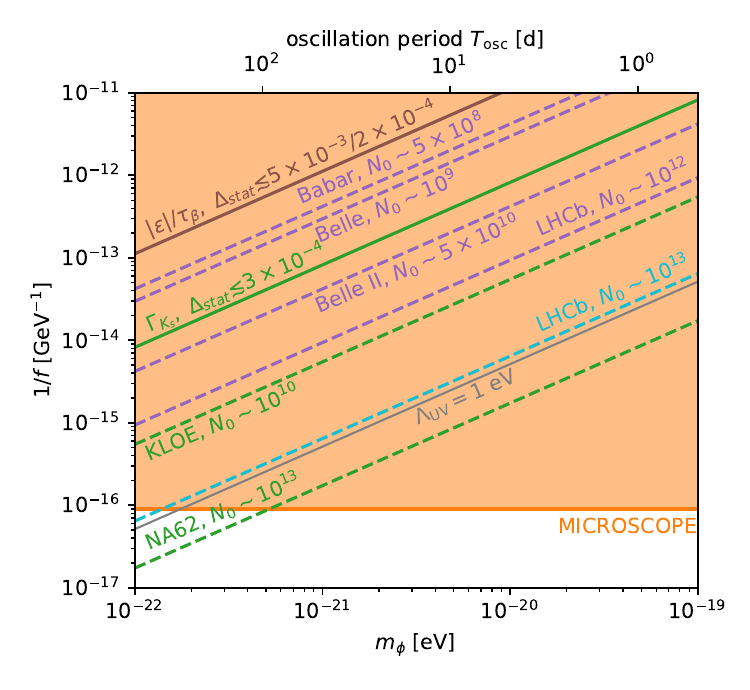}
    \caption{Potential reach of various collider searches for oscillations of the CKM matrix elements. Straight lines indicate quoted sensitivities, while dashed lines assume that for a relative measurement an $\mathcal{O}(1)$ fraction of the events that an experiment has on tape $N_0$ can be used. In green we show measurements of $|V_{us}|$ obtained from Kaons, in purple of $|V_{ub}|$ from B-Mesons and in blue of $|V_{cd}|$ from D-Mesons. The brown line refers to both oscillation of CP-violation in the Kaon system as well as lifetimes in $\beta$-decays, which happen to be of the same strength. The orange region is excluded from fith-force searches, while the gray line refers to the potential fine-tuning of the scalars mass. }
    \label{fig: zoomed limits}
\end{figure}

Given the large number of Kaons produced in experiments like KLOE, NA48 and NA62, searching for variations in $|V_{us}|$, setting the decay width of Kaons, is perhaps one of the most promising candidates. For the determination of $|V_{us}|$ e.g. the absolute branching ratio of $K^+\rightarrow \mu^+ \nu(\gamma)$ is used. It has been determined with a relative statistical error of $\sim 2\times 10^{-3}$  using $N_0\approx 10^{6}$ Kaons by the KLOE experiment \cite{ KLOE:2005xes}. For our purposes however every observable proportional to $|V_{us}|$ will work independent of whether it is suited to extract the matrix element. This makes the measurement of the $K_s$ lifetime possibly the leading candidate with a relative statistical error of $\sim 3\times 10^{-4}$ \cite{KLOE:2010yit}. To be optimistic one might consider such a study involving the $N_{0}\sim 10^{10}$ Kaons KLOE recorded, or even the $N_{0}\sim 10^{13}$ recorded by NA62. The Kaons produced by NA62 are however strongly boosted, which could further complicate the determination of absolute widths. The bounds one might obtain from these observables are shown in green in \cref{fig: limits,fig: zoomed limits}.

Similarly, $|V_{ub}|^2$ can be determined from the decay of B-mesons, although such decays are rare compared to the ones involving $V_{cb}$\,. The number of decays involving $|V_{ub}|^2$ is suppressed by $\approx|V_{ub}|^2/|V_{cb}|^2\approx 0.006$\,. However, given that Babar, Belle, Belle II and LHCb have produced $N_0\sim 5\times 10^8\,, 10^9\,, 5\times 10^{10}$ and $10^{12}$ $b\overline{b}$-pairs sensitivity to unconstrained parameter space might still be obtained. This can be seen from \cref{fig: zoomed limits}, where we show the resulting reach in purple. Additionally these experiments have in principle an even larger number of D-mesons on tape given that the production cross-section is about an order of magnitude larger than for the B-mesons. These allow the determination of $|V_{cd}|^2$ although only a fraction $\approx|V_{cd}|^2/|V_{cs}|^2\approx 0.05$ of D-mesons decays in an appropriate way. The expected reach is shown in blue in \cref{fig: zoomed limits} for LHCb only in order to avoid clutter. It becomes clear that via this channel b-factories should be able to rival the sensitivity of NA62\footnote{We thank Frédéric Blanc and Maria Vieites Diaz for pointing this out to us.}.

Further, one may consider CP violating observables. The Kaon system proves to be the most precise, with e.g. the decay rate of $K_L\rightarrow\pi^+\pi^-$ relative to the rate of $K_L\rightarrow\pi^\pm e^\mp\nu_e$ determined with a relative statistical error of $\sim 5\times 10^{-3}$ \cite{NA48:2006jeq}. The resulting bound is shown in brown in \cref{fig: zoomed limits}. Due to CP violation being small, one has to expect this sensitivity to scale poorer than $|V_{us}|$ when considering a sample of Kaons with a fixed size $N_0$\,. 

Lastly we also want to mention the possibility of observing variations in $|V_{ud}|$ via oscillations in the lifetime of $\beta$-decays. Since $|V_{ud}|^2\approx 1-|V_{us}|^2$ the relative oscillations are suppressed by $\approx|V_{us}|^2/|V_{ud}|^2\approx\lambda_\mathrm{Cabbibo}^2$ compared to the observables discussed above. Nevertheless radioactive nuclei are abundantly available and one might therefore hope to accomplish a competitive sensitivity. The measurements used to determine $|V_{ud}|$ currently achieve relative uncertainties of $\approx 2\times 10^{-4}$ \cite{Towner:2010zz}, which taking into account the reduced sensitivity leads to a bound with about the same strength as the search for CP violation we just discussed and is shown by the same brown line in \cref{fig: zoomed limits}. For these measurements it is however imperative that the sample only consists of one isotope of a given element. In order to observe oscillations this is not required but in principle any highly radioactive sample should work. 

Further details on the oscillation of CKM elements can be found in \cref{sec:experimental implications}. There we also comment on the possibility of observing an apparent unitarity-violation of the CKM matrix when combining results from experiments that ran over different periods of time. We do not expect competitive bounds from such time averaged methods though. 

%%%%%%%%%%%%%%%%%%%%%%%%%%%%%%%%%%%%%%%%%%%%%%%%%%%%%%%%%%
\section{\label{sec:challenges}Challenges and prospects }
%%%%%%%%%%%%%%%%%%%%%%%%%%%%%%%%%%%%%%%%%%%%%%%%%%%%%%%%%%

\begin{figure}
\centering
\begin{tikzpicture}
  \begin{feynman}
    \vertex (a1);
    \vertex[right=1cm of a1] (a2);
    \vertex[right=2cm of a2] (a3);
    \vertex[right=1cm of a3] (a4);
    \vertex[right=2cm of a4] (a5);

    \vertex[below=2em of a1] (b1);
    \vertex[right=1cm of b1] (b2);
    \vertex[right=2cm of b2] (b3);
    \vertex[right=1cm of b3] (b4);
    \vertex[right=2cm of b4] (b5);

    \vertex[below=0.5em of a1] (l0){\(\ \)};
    \vertex[right=0.5cm of l0] (l1){\(V(\phi)\)};
    \vertex[right=2cm of l1] (l2){\(V^\dagger(\phi)\)};
    \vertex[below=0.5em of a5] (l3){\(\ \)};
    \vertex[right=-0.5cm of l3] (l4){\(m_{q}(\phi)\)};
    
    \diagram* {
      {[edges=fermion]
        (a1) -- [edge label=\(u\)] (a2) --[edge label=\(d\)] (a3) -- [edge label=\(u\)](a4)
      },
      (a2) -- [boson,half left, edge label=\(W\)] (a3),

      (a2) -- [scalar] (b2),
      (a3) -- [scalar] (b3),
      (a5) -- [scalar] (b5)
    };
    \draw (a5) arc [start angle=270, end angle=-90, radius=0.5cm][fermion];
  \end{feynman}
\end{tikzpicture}
\caption{Left: Contribution to the up-type quark selfenergy leading to a dependence of the quark masses on $\phi$\,. Right: Correction to the scalar potential induced by quark mass dependence.}\label{fig: selfenergy}
\end{figure}
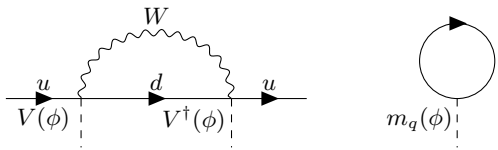

The variation of the absolute values of the CKM matrix in \cref{eq: CKM matrix dependence phi} leads to a dependence on $\phi$ of the quark masses through quantum corrections. Such couplings of $\phi$ to the quark masses are strongly constrained by searches for violations of the equivalence principle. In the near future, one can expect even stronger bounds from nuclear clocks. The coupling  to the quark masses further generates a mass correction for $\phi$ itself. This is used to judge whether the small masses considered in the previous section are fine tuned.

To get an intuition of how the coupling of $\phi$ to the quark masses arises, it is instructive to consider the contribution to the quarks self-energy depicted in \cref{fig: selfenergy}. To quantify the effect we however consider the running above the electroweak scale of the Yukawa couplings given in \cref{eq: effective quark mass matrix}, $\tilde{y}^u=\tilde{m}_u/v$\,. The field $\phi$ sets the Yukawa couplings at a high scale $\Lambda_{\rm UV}$\,. With the set of parameters chosen in \cref{sec:model}, the eigenvalues of the Yukawa couplings at this high scale are independent of $\phi$\,, but the effective CKM matrix from \cref{eq:CKM matrix} does depend on $\phi$\,. In \cref{sec:coupling to quark masses} we find that the induced dependence on $\phi$ of the eigenvalues of the Yukawa matrices at the low scale is given by 
\begin{align}
    \frac{\Delta \tilde{m}_{i}(\phi)}{\tilde{m}_{i}}&=- \frac{3}{32\pi^2}\sum_{j}|V_{ij}(\phi)|^2\tilde{y}^2_{j}\log\left(\frac{\Lambda_{\text{UV}}}{v}\right)\nonumber\\
    \frac{\Delta \tilde{m}_{j}(\phi)}{\tilde{m}_{j}}&=- \frac{3}{32\pi^2}\sum_{i}|V_{ij}(\phi)|^2\tilde{y}^2_{i}\log\left(\frac{\Lambda_{\text{UV}}}{v}\right)\,.\label{eq: quark mass variation}
\end{align}
We expect additional contributions from running below the electroweak scale to be of the same order as the result above. To get conservative estimates of the effects caused by this coupling, we take $\log(\Lambda_{\text{UV}}/v)\sim 1$ in the following.

\begin{figure}
    \centering
    \includegraphics[width=.47\textwidth]{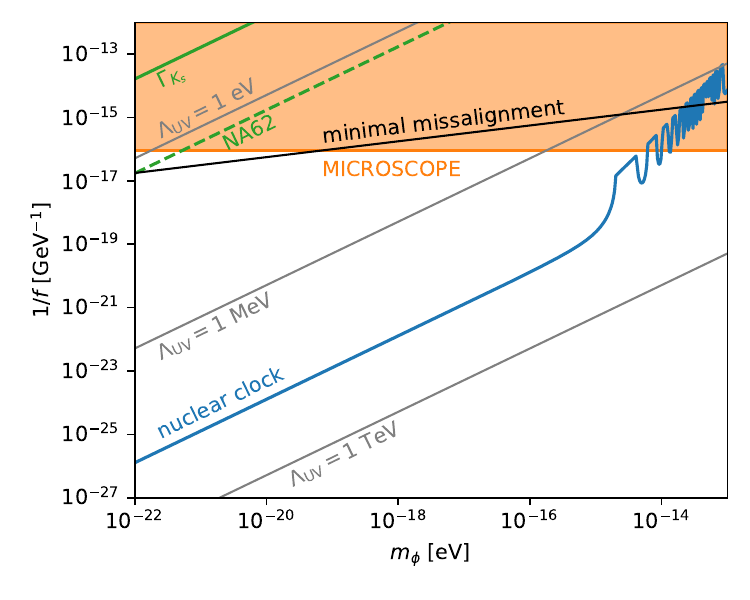}
    \caption{Bounds on the new light scalar in terms of its mass and decay constant. The orange area is excluded from searches for violation of the equivalence principle. The area above the green lines can potentially be probed through oscillations in the CKM matrix. See \cref{fig: zoomed limits} for details. Above the blue line the model can be tested by the nuclear clock. The gray lines indicate the cut-off for which the mass of the scalar is naturally small. The black line correspond to the minimal misalignment UDM model, see text.}
    \label{fig: limits}
\end{figure}

The strongest current bounds on such couplings stem from the MICROSCOPE mission \cite{Touboul:2017grn,MICROSCOPE:2022doy} searching for violations of the equivalence principle \cite{Banerjee:2022sqg}. 
MICROSCOPE constraints the differential acceleration between a platinum and a titanium test mass relative to the common gravitational acceleration caused by earth to be less than $\eta=(-1.5\pm2.7)\times 10^{-15}$\,\cite{MICROSCOPE:2022doy}. 
Using \cref{eq: quark mass variation} one can easily verify that the largest variations in mass occur for the down and the strange quark with the charm in the loop.
We follow \cite{Damour:2010rp} to estimate the resulting variations of atomic masses, where for the strange quark we only take into account the variation of the nucleon masses \cite{Junnarkar:2013ac,Shifman:1978zn}.
In this way we arrive at a bound $1/f<9\times 10^{-17}$~GeV$^{-1}$ for masses $m_{\phi}<10^{-13}$~eV, shown in orange in \cref{fig: limits,fig: zoomed limits}.
From \cref{eq: quark mass variation} it also becomes clear why we constructed the model in such a way that the CKM matrix elements involving the top quark do not depend on $\phi$\,. The large top Yukawa would lead to much stronger bounds stemming from the variation of the down type quark masses.

In the future, nuclear clocks are expected to drastically improve the sensitivity to variations in the quark masses~\cite{Campbell:2012zzb,Peik:2020cwm,PhysRevLett.132.182501,Elwell:2024qyh}. Through the coupling discussed above our model would introduce periodic oscillations in the quark masses, just like in the CKM entries, if the scalar constitutes dark matter. In \cref{fig: limits} we have indicated the expected reach in blue assuming that the scalar is all of dark matter and using the clock parameters from \cite{Banerjee:2022sqg,Banerjee:2020kww}.

Note that in our discussion of the limits from the coupling to matter we have linearized the full dependence of $\Delta m_q$ on $\phi$ in \cref{eq: quark mass variation}. In principle the scalar also has quadratic and higher order couplings. Quadratic couplings of scalars have received a large amount of attention recently, since they introduce the phenomena of screening by earths density \cite{Hees:2018fpg,Banerjee:2022sqg} as well as new signatures in experiments \cite{Masia-Roig:2022net,Flambaum:2023bnw,Kim:2023pvt}. In our case the linear couplings however dominate the prospects of detection. Further we have ensured that screening effects are negligible in the considered parameter space, as a rough estimate gives $1/f_\text{crit}\sim 10^{-9}~\text{GeV}^{-1}$ for the critical coupling \cite{Banerjee:2022sqg}.

A coupling of the type $\mathcal{L}\supset \Delta m_q(\phi)\overline{q}q$ additionally introduces a correction to the scalar potential of $\phi$ through a tadpole shown in \cref{fig: selfenergy}
\begin{equation}
    V(\phi)\supset \Delta m_q(\phi)\frac{m_q\Lambda^2_{\text{UV}}}{16\pi^2}\,,
\end{equation}
where $\Lambda_{UV}$ is a UV cutoff. The minima of this potential are CP conserving, as one can see from the exact expressions for $\Delta m_q(\phi)\propto |V_{UD}(\phi)|^2$ given in \cref{sec:experimental implications}, where all $\phi$-dependence is $\propto \cos{2(\theta_0+\phi/f)}$\,. This is expected since our effective Lagrangian \cref{eq:lagrangian} is CP conserving \cite{Vafa:1984xg}. The part of the potential that generates the CP violating VEV $\theta_0$ is therefore necessarily due to additional new physics. The term above gives a correction to the scalar mass, $\Delta m_{\phi}$\,, which is dominated by the charm and bottom quark  contributions:
\begin{equation}
    \Delta m_{\phi}\simeq \frac{\sqrt{12}}{16\pi^2}|V_{cb}||V_{ub}|\sin{2\theta_{12}}\cos{2\theta_0}\ y_c y_b\frac{v \Lambda_{\text{UV}}}{f}\,. 
\end{equation}
A model where $m_\phi\ll\Delta m_\phi$ would be considered fine tuned. Saturating $m_\phi\sim \Delta m_\phi$ we can in fact work out the amplitude of oscillation for a given UV cut-off as in \cref{eq: amplitude observables}
\begin{equation}
    \frac{\Delta O}{\overline{O}}\sim 1.1 \times 10^{-6}\frac{\text{eV}}{\Lambda_{\mathrm{UV}}}\,.
\end{equation}
In \cref{fig: limits,fig: zoomed limits} we have indicated the maximal cutoff in gray.
We conclude that, while corrections to the scalar mass are suppressed by the smallness of the Yukawa couplings, by the smallness of the off-diagonal entries of the CKM matrix, and by only arising at two loops, the amplitudes that we highlighted as detectable in the previous section still require a decent amount of fine-tuning. The mirror models discussed in Refs.~\cite{Chacko:2005pe,Burdman:2006tz,Craig:2014aea} can account for cut-offs as low as the MeV scale through the introduction of a hidden QCD sector.
As shown in~\cref{fig: limits}
nuclear clocks will be able to probe both natural regions and regions consistent with a minimal misalignment scenario for UDM production \cite{Preskill:1982cy}, satisfying the relation $f\gtrsim 10^{18}\rm\, GeV\,\left({10^{-27} \,eV\over m_\phi}\right)^{1/4},$ indicated by the black line.

Let us finally remark on corrections introduced by Planck suppressed operators, that may lead to a quality problem~\cite{Kamionkowski:1992mf,Randall:1992ut,Holman:1992us,Barr:1992qq,Perez:2020dbw,Davidi:2017gir,Vecchi:2014hpa,Banerjee:2022wzk}. Consider corrections to the scalar potential of the form 
\begin{equation}
    {\cal L} \supset c \frac{\Phi^n}{m_{\rm Pl}^{n-4}}+{\rm h.c.}\,,
\end{equation}
where $m_{\rm Pl}=2.4\times10^{18}$~GeV is the Planck mass and $c=\mathcal{O}(1)$ is a complex coefficient. Such operators lead to a correction of the scalars mass 
\begin{equation}\label{eq:massM}
    \Delta m_{\phi}\simeq n \frac{f^{n/2-1}}{m_{\rm Pl}^{n/2-2}}\,.
\end{equation}
For $\Delta m_\phi\lesssim 10^{-20}$~eV and $f=10^{16}$~GeV this implies that operators with $n\lesssim 40\,$ need to be forbidden. Apart from corrections to the mass, one needs to worry about additional contributions to the strong CP phase from these operators. If we were to consider the QCD axion $\phi\rightarrow a$ the above modification of the potential would introduce such a correction leading to the following bound,
\begin{equation}\label{eq:QCDaxion}
    \Delta m_{a}\lesssim n \frac{f^{n/2-1}}{\sqrt{\theta_{\rm QCD}}m_{\rm Pl}^{n/2-2}}\,.
\end{equation}
We notice that as the parametric dependence of \cref{eq:QCDaxion} is the same as that of \cref{eq:massM}, the bound is more stringent. One can therefore say that the Nelson-Barr dark matter framework is of higher quality.

However, in the Nelson-Barr framework there are additional Planck suppressed operators that potentially spoil the structure of the quark masses discussed in \cref{sec:model}.
For instance (see \cite{Asadi:2022vys,Perez:2020dbw} for more discussion) an operator of the form 
\begin{equation}
    {\cal L} \supset c'\frac{\Phi^n}{m^{n-1}_{\rm Pl}}\overline{q}q +{\rm h.c.}\,
\end{equation}
will contribute to the strong CP phase. 
However, unlike the case of the PQ-symmetry of the QCD-axion, here we rely on a non-anomalous U(1) flavor symmetry. Therefore, the additional dangerous operators can be sufficiently suppressed by gauging a $\mathds{Z}_n$ subgroup of the flavor symmetry. We give a complete discussion for a particular UV completion in \cref{sec:model building}. In conclusion, it seems that the quality problem of our Nelson-Barr DM construction is significantly ameliorated compared to the QCD-axion case.

%%%%%%%%%%%%%%%%%%%%%%%%%%%%%%%%%%%%%%%%%%%%%%%%%%%%%%%%%%%%%%%%%%%%%%
\begin{acknowledgments}
The authors specially thank Y. Nir who contributed significantly to this project. GP thanks Y. Soreq and S. Rajendran for discussions and comments.

MD is supported in part by U.S. Department of Energy grant No. DE-FG02-04ER41286.
The work of GP is supported by grants from BSF-NSF, Friedrich Wilhelm Bessel research award of the Alexander von Humboldt Foundation, GIF, ISF, Minerva,
SABRA - Yeda-Sela - WRC Program, the Estate of Emile Mimran, and the Maurice and Vivienne Wohl Endowment.
IS is supported by the Office of High Energy Physics of the U.S. Department of Energy under contract DE-AC02-05CH11231 and by the CHE/PBC Fellowship for Outstanding Women Postdoctoral Fellows.
\end{acknowledgments}

\section*{Supplemental Material}
\appendix

%%%%%%%%%%%%%%%%%%%%%%%%%%%%%%%%%%%%%%%%%%%%%%%%%%%%%%%%%%%%%%%%%%%%%%
\section{Flavor structure from horizontal $U(1)$ symmetries}
\label{sec:model building}
%%%%%%%%%%%%%%%%%%%%%%%%%%%%%%%%%%%%%%%%%%%%%%%%%%%%%%%%%%%%%%%%%%%%%%
To see how the special structure of coupling constants discussed in the main text might arise, here we consider two realizations under which the $U(1)$ symmetry $\theta\rightarrow\theta+\Delta \theta$ is exact and anomaly-free, as well as the necessary breaking pattern. One possible charge assignment containing the fields of the Lagrangian \cref{eq:lagrangian} is
\begin{equation}
\begin{split}
    Q^{U(1)}&(\Phi,\ u_1, Q_1, d_1,\ u_2, Q_2, d_2)=\\
&(+1,\ +1,+1,+1, \ -1,-1,-1)\,,
\end{split}
\end{equation}
while all other fields are neutral. For the symmetry to be realized exactly one needs $g\propto(1,0,0),\ \tilde{g}\propto (0,1,0)$ as well as the Yukawa matrices to be diagonal. 

It is clear that in order to get to the model discussed in the main text, off-diagonal entries in the Yukawa matrices must break the symmetry. In general the real Yukawa matrices appearing in \cref{eq:lagrangian} can be decomposed as $y^u=O^{u,T}_L \text{diag}(y_u,y_c,y_t) O_R^u~,$ $y^d=O^{d,T}_L \text{diag}(y_d,y_s,y_b) O_R^d~,$ where the rotations $O\in SO(3)$\,. Just like in the SM $O_R^d$ can be removed and only $V^{d}=O_{L}^u O_{R}^{d,T}$ is physical due to the invariance of the quark kinetic terms under these rotations. Contrary to the SM, $O_{R}^u$ cannot be removed if one wants to keep $g\propto(1,0,0),\ \tilde{g}\propto (0,1,0)$ though. One can of course exchange $O_{R}^u$ for spurions in $g,~\tilde g$ that keep them orthogonal. Additionally one can consider spurions that lead to $\tilde g^\dagger g\neq 0$\,. Both $O_{R}^u$ as well as $\tilde g^\dagger g\neq 0$ lead to a dependence of the effective mass eigenvalues $\tilde m_u$ from \cref{eq: effective quark mass matrix} on $\theta$ at tree-level though. This can be seen by considering the product of the eigenvalues
\begin{align}
    \det\left(\tilde m_u \tilde m_u^\dagger\right)&=\det\left( m_u  m_u^\dagger\right)\det\left(\mathbb{1}-\frac{B B^\dagger}{\mu^2+|B|^2}\right)\\
    &=\det\left( m_u  m_u^\dagger\right)\left(1-\frac{|B|^2}{\mu^2+|B|^2}\right)\\
    |B|^2&=\frac{f^2}{2}\left(|g|^2+|\tilde g|^2+2|\tilde g^\dagger g|\cos(2\theta)\right)\,.
\end{align}
It is therefore clear that $|\tilde g^\dagger g|=0$ is required for the mass eigenvalues to be independent of $\theta$\,. Similarly one finds for the sum of the eigenvalues
\begin{align}
    \mathrm{tr}\left(\tilde m_u \tilde m_u^\dagger\right)&=\mathrm{tr}\left( m_u  m_u^\dagger\right)-\mathrm{tr}\left(\frac{B B^\dagger}{\mu^2+|B|^2} m_u^\dagger m_u\right)\\
    &=\mathrm{tr}\left( m_u  m_u^\dagger\right)-\frac{1}{\mu^2+|B|^2} B^\dagger m_u^\dagger m_u B\,.
\end{align}
The expression above has a $\theta$ dependence if $\tilde g^\dagger m_u^\dagger m_u g\neq 0$\,. Keeping $g\propto(1,0,0),\ \tilde{g}\propto (0,1,0)$\,, in general $O_{R}^u$ introduces such a dependence. Exceptions are rotations that only mix the first and third or second and third generation. This is however expected since one can charge the third generation $+1$ or $-1$ in order to make these rotations not brake the symmetry.

A symmetry breaking pattern that only introduces $V^{d}$ might for example be realized by further symmetries allowing for real spurions in the off-diagonal entries of $y^d$ only. The alignment between $g,\tilde g$ and the eigenvectors of $y_u^\dagger y_u$ would thus be kept, avoiding a dependence of the eigenvalues of $\tilde m_u$ on $\theta$ at tree-level.
Note further that in such a model we do not need to require an additional $\mathds{Z}_2$ symmetry to forbid operators proportional to $\tilde H Q_{1/2}\overline q$ and $\Phi q\overline q$ that potentially spoil the Nelson Barr mechanism by reintroducing a strong CP phase. The operator $\tilde H Q_{3}\overline q$ introduces no strong CP phase as long as $g_3,\tilde{g}_3=0$\,.

Regarding the quality problem, in particular the introduction of a strong CP phase by higher dimensional operators, note that for any combination of operators constructed from the fields in \cref{eq:lagrangian} respecting the $U(1)$ symmetry the phase of $\Phi$ is unobservable by SM physics. Any observable effect therefore must be related to the additional spurions. 
Let us for concreteness assume that the third generation of quarks carries charge $n+1$ (so far it was a free parameter) and that the most off-diagonal entries of  $y^d$ are given by spurions of the form $(\eta/\Lambda),\ (\eta^{*}/\Lambda)$\,, where $\eta$ is an additional scalar field carrying charge $n$ under the $U(1)$ (see the model below for more details). The real vev of $\eta$, additionally breaking the $U(1)$, might be much smaller than $f$ such that we don't need to worry about operators suppressed by $\eta/m_{\rm Pl}$\,. The leading contribution to the CP phase is then due to
\begin{align}
    {\cal L} \supset c \frac{\Phi^{*n}}{m_{\rm Pl}^{n}}H Q_3 \bar d_{1}+{\rm h.c.}\,.
\end{align}
Interference with the off-diagonal term generated by $\eta/\Lambda_{\rm spur}$ gives a strong CP phase $\sim c|V_{ub}|(f/m_{\rm Pl})^n$. When only gauging a $\mathds{Z}_n$ subgroup of the $U(1)$, we therefore need to require that $n\gtrsim 4$ for $f\sim 10^{16}$~GeV in order to comply with the current upper bound on the strong CP phase. \\

\begin{figure}[h]
\includegraphics[width=6cm]{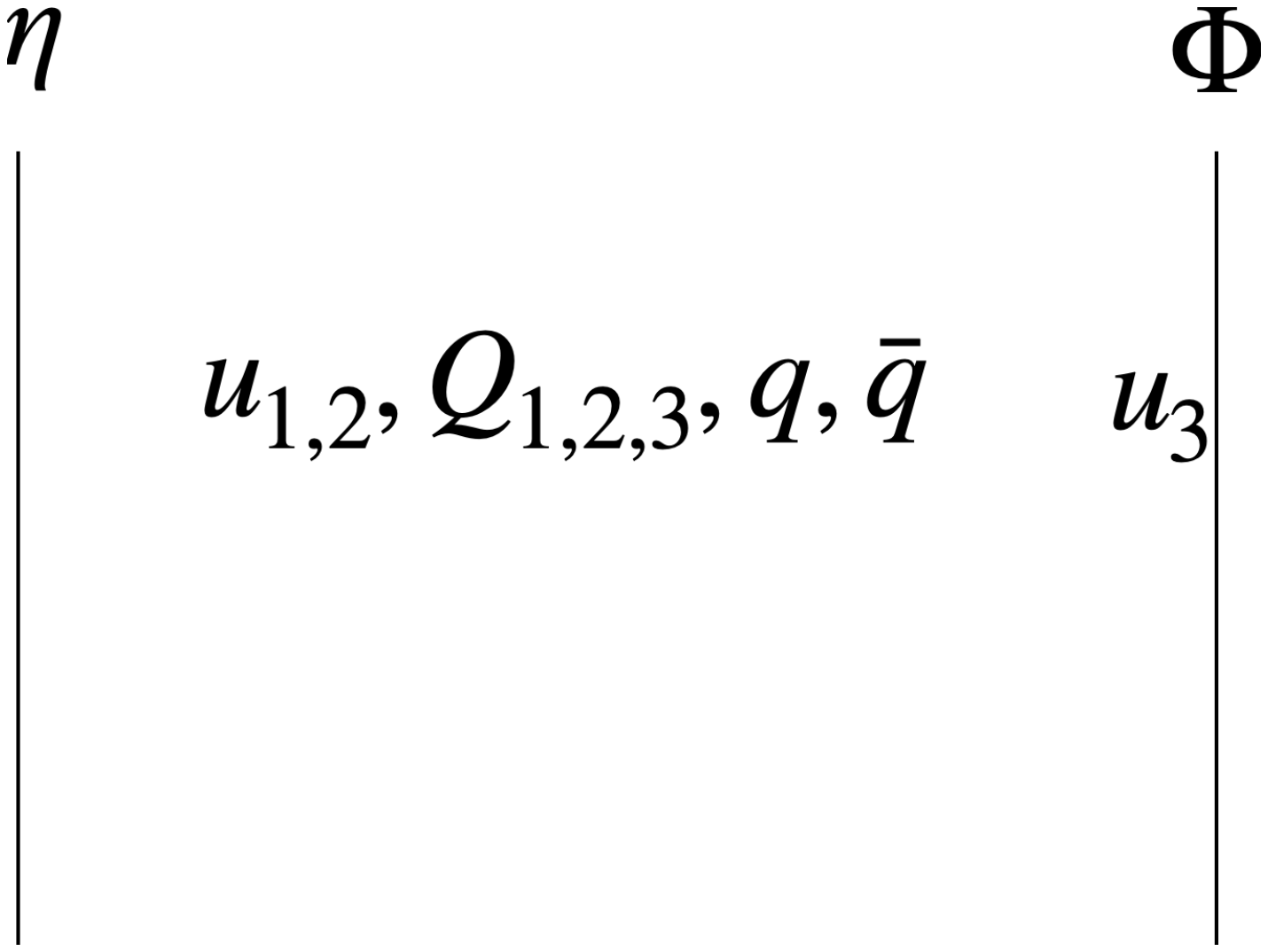}
\caption{Field content of the second model, with the SM fields mainly living in the bulk, while the two symmetry breaking scalars live on separate branes.}
\label{fig: GiladsModel}
\end{figure}

For the second model we  add a scalar $\eta$ (in addition to the field content of \cref{eq:lagrangian}), and again impose a non-anomalous $U(1)$ symmetry under which the fields carry the following charges:
\begin{equation}
Q^{U(1)}(\eta,\Phi,q,\bar q,\bar u_1)=(+1,+1/2,-1/2,-1/2,+1)\,.
\end{equation}
The $U(1)$ symmetry is spontaneously broken by the vevs of both $\eta$ and $\Phi$\,. Those are sequestered from each other, e.g. living on different branes with the SM fields living in the bulk, see \cref{fig: GiladsModel}. Furthermore, on the $\eta$-brane there is a source of soft breaking such that the axionic component of $\eta$ receives a mass. It is therefore not relevant to the rest of the discussion.

Under these assumptions, the effective Lagrangian is given by
\begin{equation}
    {\cal L} \supset \langle\eta\rangle \bar q q + (g \Phi \bar u_2+\tilde g\Phi^*\bar u_1)  q+  y^{u}_{ij} \tilde{H}Q_i \bar u_{j}+ y^{d}_{ij} H Q_i \bar d_{j}+ \dots
\end{equation}
where $y^{u}_{ij}$ is the same as $y^{u}_{ij}$ in the original Lagrangian \cref{eq:lagrangian} except for the elements involving $\overline{u}_1$\,, $y^{u}_{i1}\to \epsilon y^{u}_{i1}$ 
with $\epsilon=\langle\eta^*\rangle/\Lambda$\,, where $\Lambda$ is some UV scale above the VEV of $\eta$\,. The $\eta$ therefore provides the additional breaking of the shift-symmetry to ensure that the VEV of $\theta$ breaks CP as required to account for the observed value of the CKM phase.

In this model the mass scale $\mu=\langle\eta\rangle$ is not related to $f$\,, which offers an explanation for why $\mu\ll|B|$\,. This is required to get the observed quark mass hierarchy as well as $\theta_{12}\sim\lambda_{\rm Cabbibo}$ as explained in \cref{sec:experimental implications}. In models where $\mu$ and $f$ are related to the same UV scale, this potentially poses a problem since $|g|,|\tilde g|\lesssim10^{-3}$ is required in order for loop contributions to the strong CP phase to be sufficiently suppressed \cite{Dine:2015jga}. Note that the model further directly offers an explanation for why the Yukawas involving the up-quark are small. One can therefore hope to complete a construction along these lines into a complete flavor model. Additionally in this model the operators $\tilde H Q\overline q$ and $\Phi q\overline q$ are already forbidden by the $U(1)$ symmetry and the $\mathds{Z}_2$ symmetry is not required. 

A draw-back of this second model compared to the first one is that it does not as easily explain the special structure of couplings that we introduced in \cref{sec:model}, that ultimately avoided the dependence of the quark masses at tree-level as well as suppressing the loop contributions. There is no reason $g \Phi \bar u_3 q$ vanishes, leading to larger loop contributions due to the large top Yukawa. This issue however might still be solved by placing $u_3$ mostly on the $\Phi$-brane  with the first and second generations allowed to be in the bulk (see \cref{fig: GiladsModel}). This idea has been pursued previously in flavor models. Further the model provides no explanation for the absence of $O^u_R$ that leads to a dependence of the quark masses on the new scalar at tree-level.

%%%%%%%%%%%%%%%%%%%%%%%%%%%%%%%%%%%%%%%%%%%%%%%%%%%%%%%%%%%%%%%%%
\section{Details of the coupling to the CKM matrix}
\label{sec:experimental implications}
%%%%%%%%%%%%%%%%%%%%%%%%%%%%%%%%%%%%%%%%%%%%%%%%%%%%%%%%%%%%%%%%%
Given our construction \cref{sec:model}, the CKM elements become
\begin{align}
V_{uj}&=V^{d}_{uj}\left(\cos{{\theta}_{12}}+\frac{V^{d}_{cj}}{V^{d}_{uj}}\sin{{\theta}_{12}}\exp\left[2i\left(\theta_0+\frac{\phi}{f}\right)\right]\right)\,,\\
V_{cj}&=V^{d}_{cj}\left(\cos{{\theta}_{12}}\exp\left[2i\left(\theta_0+\frac{\phi}{f}\right)\right]-\frac{V^{d}_{uj}}{V^{d}_{cj}}\sin{{\theta}_{12}}\right)\,,
\end{align}
where $V^{d}$ are the CKM entries coming from the misalignment of the $y^u$ and $y^d$ matrices only, and ${\theta}_{12}$ is the rotation angle required to diagonalize $\tilde{m}_u{\tilde{m}_u}^\dagger$\,, given by
\begin{equation}
    \sin{2{\theta}_{12}}=\frac{2|g||\tilde{g}|~a^2 m_u m_c}{\tilde{m}^2_u-\tilde{m}^2_c}\,,\ a^2 = \frac{f^2}{2\mu^2+f^2\leri{|g|^2+ |\tilde{g}|^2}}\,.
\end{equation}
The up and the charm masses are given by
\begin{equation}
    2\tilde{m}^2_{c,u}=M^2\pm\sqrt{M^2+4(a^2(|g|^2+|\tilde g|^2)-1)m_u^2m_c^2}\,,
\end{equation}
with 
$M^2=(1-a^2 |g|^2)m_u^2+(1-a^2 |\tilde{g}|^2)m_c^2$\,.
For simplicity we consider $|g|=|\tilde{g}|$ from now on.
We define $\epsilon$ via $|g|^2a^2=\frac12\leri{1-\epsilon^2}$\,, with $\epsilon^2\approx \frac{\mu^2}{|g|^2f^2}$\,. If $\epsilon \ll 1$\,, we obtain
\begin{align}
\tilde{m}^2_{u}&\approx 2\epsilon^2 \frac{m_u^2m_c^2}{m_u^2+m_c^2}\,,\\
\tilde{m}^2_{c}&\approx \frac{1}{2}\leri{m_u^2+m_c^2}\,,\\
\sin{2{\theta}_{12}}&\approx-\frac{2m_u m_c}{m_u^2+m_c^2}\approx-\frac{1}{\epsilon}\frac{\tilde{m}_u}{\tilde{m}_c}\,.
\end{align}
Note that since we require $\theta_{12}\gtrsim\lambda_{\rm Cabbibo}$ in order to achieve a large enough CP violation the original hierarchy between $m_u$ and $m_c$ is rather small. The observed large hierarchy is therefore necessarily due to $\epsilon\ll 1$ in this model.

CP-violating observables are proportional to the Jarlskog invariant
\begin{align}
    J &=  \text{Im}\leri{V_{ud}V^*_{ub}V_{tb}V^*_{td}}\nonumber\\
    &= \frac{1}{2}|V_{tb}||V_{td}| \sin{2\theta}\sin{2{\theta}_{12}}\leri{V^{d}_{cd}V^{d}_{ub}-V_{cb}^{d}V^{d}_{ud}}\,.
\end{align}

The absolute values $|V_{uj}|^2$ and $|V_{cj}|^2$ become
\begin{align}
|V_{u/cj}|^2&=|V^{d}_{u/cj}\cos{{\theta}_{12}}|^2\times\Bigg(1+\left(\frac{V^{d}_{c/uj}}{V^{d}_{u/cj}}\right)^2\tan^2{{\theta}_{12}}\nonumber\\
&\qquad\pm2\frac{V^{d}_{c/uj}}{V^{d}_{u/cj}}\tan{{\theta}_{12}}\cos\leri{2\theta_0+2\frac{\phi}{f}}\Bigg)\,.
\end{align}
Decomposing into $\phi$-dependent and $\phi$-independent terms, in the limit of $\phi/f\ll 1$\,, we get
\begin{align}
|V_{u/cj}|^2&\approx\mp2V^{d}_{u/cj}V^{d}_{c/uj}\kappa\phi/f+|V^{d}_{u/cj}|^2\times\nonumber\\
&\times
\left(\cos^2{{\theta}_{12}}+\left(\frac{V^{d}_{c/uj}}{V^{d}_{u/cj}}\right)^2\sin^2{{\theta}_{12}}+\frac{V^{d}_{c/uj}}{V^{d}_{u/cj}}\kappa\right)\,.
\end{align}
Note that the assumption $|V_{ij}^d|\approx |V_{ij}|$ is valid as long as $\frac{V^{d}_{u/cj}}{V^{d}_{c/uj}}\tan{{\theta}_{12}}\lesssim1$\,. This implies $\theta_{12}\sim\lambda_{\rm Cabbibo}$ and an $\mathcal{O}(1)$ value for $\theta_0$ in order to match the observed CP-violation. This justifies our choice in the main text. 

\begin{figure}
    \centering
    \includegraphics[width=.45\textwidth]{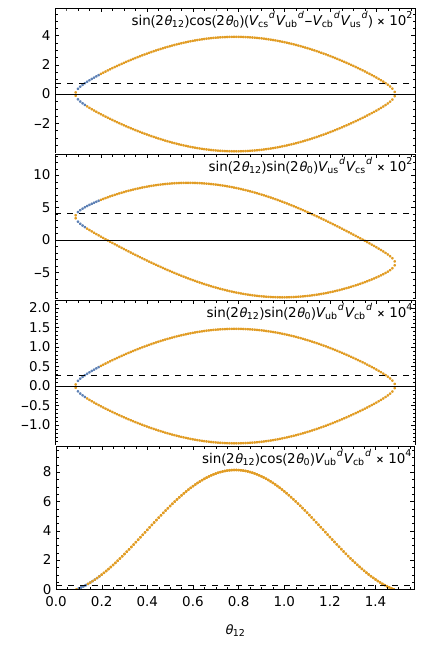}
    \caption{Scan over the parameter $\theta_{12}$\,, choosing $\theta_0,~V^{d}$ such that the time independent part matches the observed CKM matrix. The shown combinations from top to botton are relavant for the linear dependence of the Jarlskog, $|V_{us}|$ and $|V_{ub}|$ as well as for the quadratic dependence of $\Delta m_{b}$ on $\phi$\,. The last one dominates the correction of the scalar mass through the tadpole. For the blue points $|V_{UB}|\approx|V^{d}_{UB}|$ is satisfied up to a factor of 2 for all matrix elements. The dashed lines give the value we used to estimate the reach of experiments in the main text.}
    \label{fig: parameter scan}
\end{figure}

Going beyond this assumption we scan in \cref{fig: parameter scan} over arbitrary values of $\theta_{12}$\,. Depending on $\theta_{12}$ the value of $\theta_0$ and $V^{d}$ are chosen such that the time independent part of the CKM matrix matches the observed one. Note that there is a minimal $\theta_{12}$ for which such solutions exist, since for $\theta_{12}\rightarrow0$ the observed CP-violation cannot be matched. We show the combinations of the parameters that are relevant for the linear dependence of CP-violating observables, $|V_{us}|$ and $|V_{ub}|$\,. The bottom plot gives the combination that dominates the scalars mass correction. For the blue points $|V_{ij}^y|\approx |V_{ij}|$ is fulfilled for all matrix elements up to a factor of 2. We see that for those points the actual value matches up to an $\mathcal{O}(1)$ factor the approximations used in the main text, shown by the dashed lines.

%%%%%%%%%%%%%%%%%%%%%%%%%%%%%%%%%%%%%%%%%%%%%%%%%%%%%%%%%%%%%%%%%%%%
\section{Time-averaged Analysis}
%%%%%%%%%%%%%%%%%%%%%%%%%%%%%%%%%%%%%%%%%%%%%%%%%%%%%%%%%%%%%%%%%%%%
Without time-stamping the events, we expect effects linear in $\phi/f$ to be averaged out if $m_\phi>1/\tau_{\rm exp}$ with $\tau_{\rm exp}$ being the total experimental duration, following
\begin{align}
\Delta_1&|V_{uj}|^2\approx\nonumber\\
&-\frac{2V^{d}_{cj}V^{d}_{uj}\kappa\phi_0 }{f}\sinc{\frac{m_\phi \tau_{\rm exp} }{2}} \sin \left(\alpha +\frac{m_\phi \tau_{\rm exp} }{2}\right)\,,
\end{align}
where $\text{sinc}\leri{x}\equiv\sin{x}/x$\,, $\phi_0$ is the oscillations amplitude $\phi_0=\sqrt{2\rho_{\rm DM}}/m_\phi$ and $\alpha$ the phase of the cosine in Eq.~\eqref{eq:uldm_osc} at the onset of the experiment.
One may also consider the quadratic shift
\begin{align}
\Delta_2|V_{uj}|^2&\approx -V^{d}_{cj}V^{d}_{uj}\sin{2{\theta}_{12}}\cos{2\theta_0}\frac{\phi_0^2}{f^2 }\times\nonumber\\
&\leri{1- \sinc{m_\phi \tau_{\rm exp} } \cos \leri{2 \alpha +m_\phi \tau_{\rm exp} }}\,.
\end{align}

Interestingly, if $|V_{uj}|^2$ and $|V_{cj}|^2$ are measured at different times, and/or by experiments of different duration, a deviation from unitarity might be observed.

In the linear approximation the apparent deviation from unitarity for the $j$th down-quark is
\begin{align}
    &1-\leri{|V_{uj}|^2+|V_{cj}|^2+|V_{tj}|^2}=\frac{2V^{d}_{cj}V^{d}_{uj}\kappa\phi_0 }{f}\times\nonumber\\
    &\Bigg[\sinc{\frac{m_\phi \tau^{uj}_{\rm exp} }{2}} \sin \left(\alpha^{uj} +\frac{m_\phi \tau^{uj}_{\rm exp} }{2}\right)-\nonumber\\
    &-\sinc{\frac{m_\phi \tau^{cj}_{\rm exp} }{2}} \sin \left(\alpha^{cj} +\frac{m_\phi \tau^{cj}_{\rm exp} }{2}\right)\Bigg]\,.
\end{align}
Considering the first row of the CKM matrix, where the deviation from unitarity is $\lesssim2.5\times10^{-3}$\, at $2\sigma$\,, 
this would at most set a bound of $\phi_0/f\lesssim 10^{-2}$\,, for masses that allow an $\mathcal{O}\leri{1}$ difference between the two time-averages. For this estimate we again assumed $|V|\approx |V^d|$\,.

%%%%%%%%%%%%%%%%%%%%%%%%%%%%%%%%%%%%%%%%%%%%%%%%%%%%%%%%%%%%%%%%%%%%%%
\section{Induced coupling to quark masses}
\label{sec:coupling to quark masses}
%%%%%%%%%%%%%%%%%%%%%%%%%%%%%%%%%%%%%%%%%%%%%%%%%%%%%%%%%%%%%%%%%%%%%%
In our model the observed quark masses are independent of $\theta$ at tree-level. Here we estimate the correction to this result that occurs at 1-loop due to the dependence of $|V_{ij}|^2$ on $\theta$\,. The correction is induced by diagrams like the one shown in \cref{fig: selfenergy}. To get an estimate of the size of the coupling, we only consider the renormalisation group evolution of the Yukawas above the electro-weak scale and assume that further corrections below it will be of similar magnitude. We further assume that the new quark $q$ has already been integrated out, such that we are only dealing with the effective Yukawa couplings and masses, shown in \cref{eq: effective quark mass matrix}. For convenience we drop the tilde over the effective Yukawas throughout this section.

Above the electro-weak scale the Yukawas evolve according to \cite{Grzadkowski:1987tf}
\begin{equation}
    -16\pi^2 \frac{dy_{X}}{dt}=\left[G_X \mathds{1}-T \mathds{1}-\frac{3}{2}S_X\right]y_{X}\,,
\end{equation}
where $t=\ln(\Lambda/v_{EW})$ and $y_{X}$ denotes the $3\times3$ Yukawa matrices of species $X$\,. The terms $G_X$ and $T$ are due to gauge-couplings and traces over Yukawas. Since they are proportional to $\mathds{1}$ they only lead to a rescaling of the Yukawa matrices and their eigenvalues as a whole. Importantly they are independent of $\theta$ and therefore do not introduce the dependence we are after. We will therefore drop them from now on and only consider 
\begin{equation}
    S_{u}=y_{u}y_{u}^\dagger-y_{d}y_{d}^\dagger=-S_{d}\,.
\end{equation}
We start the evolution at some scale $t_0$ and evolve by a small amount up to $t_1=t_0+\Delta t$\,. At $t_0$ the down type Yukawas can be brought into a diagonal form $y_{d,0}=y_{d,0}^\mathrm{diag}$\,, while the up type Yukawas read $y_{u,0}=V_{0}(\theta) y_{u,0}^\mathrm{diag}$\,, where $V_{0}(\theta)$ is the $\theta$-dependent CKM matrix at the starting scale given in \cref{eq:CKM matrix}. Assuming that $\Delta t$ is small the Yukawas at $t_1$ are then given by %
\begin{equation}
    y_{u/d,1}\simeq y_{u/d,0}\pm\frac{3}{32\pi^2} \Delta t (y_{u,0}y_{u,0}^\dagger-y_{d,0}y_{d,0}^\dagger)y_{u/d,0}\,.
\end{equation}
To find the new eigenvalues we need to diagonalize $y_{u/d,1}y_{u/d,1}^\dagger$\,. To linear order in $\Delta t$ we can find the corrections to the eigenvalues simply multiplying with the eigenvectors of $y_{u/d,0}y_{u/d,0}^\dagger$\,. That means the new eigenvalues are given by the diagonal entries of $y_{d,1}y_{d,1}^\dagger$ and $V_{0}^\dagger y_{u,1}y_{u,1}^\dagger V_{0}$\,. For the $i$-th eigenvalue we find%\vspace{-0.2cm}
\begin{align}
    \frac{y^2_{u,i,1}}{y^2_{u,i,0}}&=1-\frac{3\Delta t}{16\pi^2} \left(\sum_{j}|V_{ij,0}(\theta)|^2y^2_{d,j,0}-y^2_{u,i,0}\right)\\
    \frac{y^2_{d,i,1}}{y^2_{u,i,0}}&=1+\frac{3\Delta t}{16\pi^2}  \left(y^2_{d,i,0}-\sum_{j}|V_{ji,0}(\theta)|^2y^2_{u,j,0}\right)\,.
\end{align}
Taking the square root and dropping the $\theta$ independent correction we arrive at \cref{eq: quark mass variation}. 

%%%%%%%%%%%%%%%%%%%%%%%%%%%%%%%%%%%%%%%%%%%%%%%%%%%%%%%%%%%%%%%%%%%%%%%%

\bibliography{apssamp}

@article{Elwell:2024qyh,
    author = "Elwell, R. and Schneider, Christian and Jeet, Justin and Terhune, J. E. S. and Morgan, H. W. T. and Alexandrova, A. N. and Tan, H. B. Tran and Derevianko, Andrei and Hudson, Eric R.",
    title = "{Laser excitation of the $^{229}$Th nuclear isomeric transition in a solid-state host}",
    eprint = "2404.12311",
    archivePrefix = "arXiv",
    primaryClass = "physics.atom-ph",
    month = "4",
    journal= "",
    year = "2024"
}

@article{Dine:2015jga,
    author = "Dine, Michael and Draper, Patrick",
    title = "{Challenges for the Nelson-Barr Mechanism}",
    eprint = "1506.05433",
    archivePrefix = "arXiv",
    primaryClass = "hep-ph",
    doi = "10.1007/JHEP08(2015)132",
    journal = "JHEP",
    volume = "08",
    pages = "132",
    year = "2015"
}

@article{Peccei:1977hh,
    author = "Peccei, R. D. and Quinn, Helen R.",
    title = "{CP Conservation in the Presence of Instantons}",
    reportNumber = "ITP-568-STANFORD",
    doi = "10.1103/PhysRevLett.38.1440",
    journal = "Phys. Rev. Lett.",
    volume = "38",
    pages = "1440--1443",
    year = "1977"
}

@article{PhysRevLett.132.182501,
  title = {Laser Excitation of the Th-229 Nucleus},
  author = {Tiedau, J. and Okhapkin, M. V. and Zhang, K. and Thielking, J. and Zitzer, G. and Peik, E. and Schaden, F. and Pronebner, T. and Morawetz, I. and De Col, L. Toscani and Schneider, F. and Leitner, A. and Pressler, M. and Kazakov, G. A. and Beeks, K. and Sikorsky, T. and Schumm, T.},
  journal = {Phys. Rev. Lett.},
  volume = {132},
  issue = {18},
  pages = {182501},
  numpages = {6},
  year = {2024},
  month = {Apr},
  publisher = {American Physical Society},
  doi = {10.1103/PhysRevLett.132.182501},
  url = {https://link.aps.org/doi/10.1103/PhysRevLett.132.182501}
}

@article{Georgi:1978xz,
    author = "Georgi, Howard",
    title = "{A Model of Soft CP Violation}",
    reportNumber = "HUTP-78/A010",
    journal = "Hadronic J.",
    volume = "1",
    pages = "155",
    year = "1978"
}

@article{Mohapatra:1978fy,
    author = "Mohapatra, Rabindra N. and Senjanovic, G.",
    title = "{Natural Suppression of Strong p and t Noninvariance}",
    reportNumber = "CCNY-HEP-78-12",
    doi = "10.1016/0370-2693(78)90243-5",
    journal = "Phys. Lett. B",
    volume = "79",
    pages = "283--286",
    year = "1978"
}

@article{Davidi:2017gir,
    author = "Davidi, Oz and Gupta, Rick S. and Perez, Gilad and Redigolo, Diego and Shalit, Aviv",
    title = "{Nelson-Barr relaxion}",
    eprint = "1711.00858",
    archivePrefix = "arXiv",
    primaryClass = "hep-ph",
    doi = "10.1103/PhysRevD.99.035014",
    journal = "Phys. Rev. D",
    volume = "99",
    number = "3",
    pages = "035014",
    year = "2019"
}

@article{Bento:1991ez,
    author = "Bento, Luis and Branco, Gustavo C. and Parada, Paulo A.",
    title = "{A Minimal model with natural suppression of strong CP violation}",
    reportNumber = "IFM-8-91",
    doi = "10.1016/0370-2693(91)90530-4",
    journal = "Phys. Lett. B",
    volume = "267",
    pages = "95--99",
    year = "1991"
}

@article{Grzadkowski:1987tf,
    author = "Grzadkowski, B. and Lindner, M.",
    title = "{Nonlinear Evolution of Yukawa Couplings}",
    reportNumber = "MPI-PAE/PTh-4/87",
    doi = "10.1016/0370-2693(87)90458-8",
    journal = "Phys. Lett. B",
    volume = "193",
    pages = "71",
    year = "1987"
}

@article{Banerjee:2022sqg,
    author = "Banerjee, Abhishek and Perez, Gilad and Safronova, Marianna and Savoray, Inbar and Shalit, Aviv",
    title = "{The phenomenology of quadratically coupled ultra light dark matter}",
    eprint = "2211.05174",
    archivePrefix = "arXiv",
    primaryClass = "hep-ph",
    doi = "10.1007/JHEP10(2023)042",
    journal = "JHEP",
    volume = "10",
    pages = "042",
    year = "2023"
}

@article{Junnarkar:2013ac,
    author = "Junnarkar, Parikshit and Walker-Loud, Andre",
    title = "{Scalar strange content of the nucleon from lattice QCD}",
    eprint = "1301.1114",
    archivePrefix = "arXiv",
    primaryClass = "hep-lat",
    reportNumber = "NT-LBNL-13-001, UCB-NPAT-13-001, UNH-13-01",
    doi = "10.1103/PhysRevD.87.114510",
    journal = "Phys. Rev. D",
    volume = "87",
    pages = "114510",
    year = "2013"
}

@article{Losada:2023zap,
    author = "Losada, Marta and Nir, Yosef and Perez, Gilad and Savoray, Inbar and Shpilman, Yogev",
    title = "{Time dependent CP-even and CP-odd signatures of scalar ultralight dark matter in neutrino oscillations}",
    eprint = "2302.00005",
    archivePrefix = "arXiv",
    primaryClass = "hep-ph",
    doi = "10.1103/PhysRevD.108.055004",
    journal = "Phys. Rev. D",
    volume = "108",
    number = "5",
    pages = "055004",
    year = "2023"
}

@article{Touboul:2017grn,
    author = "Touboul, Pierre and others",
    title = "{MICROSCOPE Mission: First Results of a Space Test of the Equivalence Principle}",
    eprint = "1712.01176",
    archivePrefix = "arXiv",
    primaryClass = "astro-ph.IM",
    doi = "10.1103/PhysRevLett.119.231101",
    journal = "Phys. Rev. Lett.",
    volume = "119",
    number = "23",
    pages = "231101",
    year = "2017"
}

@article{Damour:2010rp,
    author = "Damour, Thibault and Donoghue, John F.",
    title = "{Equivalence Principle Violations and Couplings of a Light Dilaton}",
    eprint = "1007.2792",
    archivePrefix = "arXiv",
    primaryClass = "gr-qc",
    doi = "10.1103/PhysRevD.82.084033",
    journal = "Phys. Rev. D",
    volume = "82",
    pages = "084033",
    year = "2010"
}

@article{Shifman:1978zn,
    author = "Shifman, Mikhail A. and Vainshtein, A. I. and Zakharov, Valentin I.",
    title = "{Remarks on Higgs Boson Interactions with Nucleons}",
    reportNumber = "ITEP-22-1978",
    doi = "10.1016/0370-2693(78)90481-1",
    journal = "Phys. Lett. B",
    volume = "78",
    pages = "443--446",
    year = "1978"
}

@article{KLOE:2005xes,
    author = "Ambrosino, F. and others",
    collaboration = "KLOE",
    title = "{Measurement of the absolute branching ratio for the $K^+ \to \mu^+ \nu(\gamma)$ decay with the KLOE detector}",
    eprint = "hep-ex/0509045",
    archivePrefix = "arXiv",
    doi = "10.1016/j.physletb.2005.11.008",
    journal = "Phys. Lett. B",
    volume = "632",
    pages = "76--80",
    year = "2006"
}

@article{Hui:2021tkt,
    author = "Hui, Lam",
    title = "{Wave Dark Matter}",
    eprint = "2101.11735",
    archivePrefix = "arXiv",
    primaryClass = "astro-ph.CO",
    doi = "10.1146/annurev-astro-120920-010024",
    journal = "Ann. Rev. Astron. Astrophys.",
    volume = "59",
    pages = "247--289",
    year = "2021"
}

@article{Banerjee:2020kww,
    author = "Banerjee, Abhishek and Kim, Hyungjin and Matsedonskyi, Oleksii and Perez, Gilad and Safronova, Marianna S.",
    title = "{Probing the Relaxed Relaxion at the Luminosity and Precision Frontiers}",
    eprint = "2004.02899",
    archivePrefix = "arXiv",
    primaryClass = "hep-ph",
    doi = "10.1007/JHEP07(2020)153",
    journal = "JHEP",
    volume = "07",
    pages = "153",
    year = "2020"
}

@article{Hees:2018fpg,
    author = "Hees, Aur\'elien and Minazzoli, Olivier and Savalle, Etienne and Stadnik, Yevgeny V. and Wolf, Peter",
    title = "{Violation of the equivalence principle from light scalar dark matter}",
    eprint = "1807.04512",
    archivePrefix = "arXiv",
    primaryClass = "gr-qc",
    doi = "10.1103/PhysRevD.98.064051",
    journal = "Phys. Rev. D",
    volume = "98",
    number = "6",
    pages = "064051",
    year = "2018"
}

@article{Kim:2023pvt,
    author = "Kim, Hyungjin and Lenoci, Alessandro and Perez, Gilad and Ratzinger, Wolfram",
    title = "{Probing an ultralight QCD axion with electromagnetic quadratic interaction}",
    eprint = "2307.14962",
    archivePrefix = "arXiv",
    primaryClass = "hep-ph",
    reportNumber = "DESY-23-110",
    doi = "10.1103/PhysRevD.109.015030",
    journal = "Phys. Rev. D",
    volume = "109",
    number = "1",
    pages = "015030",
    year = "2024"
}

@article{Masia-Roig:2022net,
    author = "Masia-Roig, Hector and others",
    title = "{Intensity interferometry for ultralight bosonic dark matter detection}",
    eprint = "2202.02645",
    archivePrefix = "arXiv",
    primaryClass = "hep-ph",
    doi = "10.1103/PhysRevD.108.015003",
    journal = "Phys. Rev. D",
    volume = "108",
    number = "1",
    pages = "015003",
    year = "2023"
}

@article{Flambaum:2023bnw,
    author = "Flambaum, V. V. and Samsonov, I. B.",
    title = "{Fluctuations of atomic energy levels due to axion dark matter}",
    eprint = "2302.11167",
    archivePrefix = "arXiv",
    primaryClass = "hep-ph",
    doi = "10.1103/PhysRevD.108.075022",
    journal = "Phys. Rev. D",
    volume = "108",
    number = "7",
    pages = "075022",
    year = "2023"
}

@article{KLOE:2010yit,
    author = "Ambrosino, F. and others",
    collaboration = "KLOE",
    title = "{Precision Measurement of $K_S$ Meson Lifetime with the KLOE detector}",
    eprint = "1011.2668",
    archivePrefix = "arXiv",
    primaryClass = "hep-ex",
    doi = "10.1140/epjc/s10052-011-1604-7",
    journal = "Eur. Phys. J. C",
    volume = "71",
    pages = "1604",
    year = "2011"
}

@article{NA48:2006jeq,
    author = "Lai, A. and others",
    collaboration = "NA48",
    title = "{Measurement of the ratio Gamma(KL ---\ensuremath{>} pi+ pi-) / Gamma(KL ---\ensuremath{>} pi e nu) and extraction of the CP violation parameter |eta(+-)|}",
    eprint = "hep-ex/0611052",
    archivePrefix = "arXiv",
    reportNumber = "CERN-PH-EP-2006-034",
    doi = "10.1016/j.physletb.2006.11.071",
    journal = "Phys. Lett. B",
    volume = "645",
    pages = "26--35",
    year = "2007"
}

@article{Towner:2010zz,
    author = "Towner, I. S. and Hardy, J. C.",
    title = "{The evaluation of V(ud) and its impact on the unitarity of the Cabibbo-Kobayashi-Maskawa quark-mixing matrix}",
    doi = "10.1088/0034-4885/73/4/046301",
    journal = "Rept. Prog. Phys.",
    volume = "73",
    pages = "046301",
    year = "2010"
}

@article{Vafa:1984xg,
    author = "Vafa, Cumrun and Witten, Edward",
    title = "{Parity Conservation in QCD}",
    reportNumber = "Print-84-0549 (PRINCETON)",
    doi = "10.1103/PhysRevLett.53.535",
    journal = "Phys. Rev. Lett.",
    volume = "53",
    pages = "535",
    year = "1984"
}

@article{Chacko:2005pe,
    author = "Chacko, Z. and Goh, Hock-Seng and Harnik, Roni",
    title = "{The Twin Higgs: Natural electroweak breaking from mirror symmetry}",
    eprint = "hep-ph/0506256",
    archivePrefix = "arXiv",
    doi = "10.1103/PhysRevLett.96.231802",
    journal = "Phys. Rev. Lett.",
    volume = "96",
    pages = "231802",
    year = "2006"
}

@article{Burdman:2006tz,
    author = "Burdman, Gustavo and Chacko, Z. and Goh, Hock-Seng and Harnik, Roni",
    title = "{Folded supersymmetry and the LEP paradox}",
    eprint = "hep-ph/0609152",
    archivePrefix = "arXiv",
    reportNumber = "SLAC-PUB-12115",
    doi = "10.1088/1126-6708/2007/02/009",
    journal = "JHEP",
    volume = "02",
    pages = "009",
    year = "2007"
}

@article{Craig:2014aea,
    author = "Craig, Nathaniel and Knapen, Simon and Longhi, Pietro",
    title = "{Neutral Naturalness from Orbifold Higgs Models}",
    eprint = "1410.6808",
    archivePrefix = "arXiv",
    primaryClass = "hep-ph",
    reportNumber = "RU-NHETC-2014-15",
    doi = "10.1103/PhysRevLett.114.061803",
    journal = "Phys. Rev. Lett.",
    volume = "114",
    number = "6",
    pages = "061803",
    year = "2015"
}

@article{Losada:2021bxx,
    author = "Losada, Marta and Nir, Yosef and Perez, Gilad and Shpilman, Yogev",
    title = "{Probing scalar dark matter oscillations with neutrino oscillations}",
    eprint = "2107.10865",
    archivePrefix = "arXiv",
    primaryClass = "hep-ph",
    doi = "10.1007/JHEP04(2022)030",
    journal = "JHEP",
    volume = "04",
    pages = "030",
    year = "2022"
}

@article{Dev:2020kgz,
    author = "Dev, Abhish and Machado, Pedro A. N. and Mart\'\i{}nez-Mirav\'e, Pablo",
    title = "{Signatures of ultralight dark matter in neutrino oscillation experiments}",
    eprint = "2007.03590",
    archivePrefix = "arXiv",
    primaryClass = "hep-ph",
    reportNumber = "FERMILAB-PUB-20-260-T",
    doi = "10.1007/JHEP01(2021)094",
    journal = "JHEP",
    volume = "01",
    pages = "094",
    year = "2021"
}

@article{Barr:1984qx,
    author = "Barr, Stephen M.",
    title = "{Solving the Strong CP Problem Without the Peccei-Quinn Symmetry}",
    reportNumber = "DOE-ER-40048-08 P4",
    doi = "10.1103/PhysRevLett.53.329",
    journal = "Phys. Rev. Lett.",
    volume = "53",
    pages = "329",
    year = "1984"
}

@article{Nelson:1983zb,
    author = "Nelson, Ann E.",
    title = "{Naturally Weak CP Violation}",
    reportNumber = "HUTP-83/A080",
    doi = "10.1016/0370-2693(84)92025-2",
    journal = "Phys. Lett. B",
    volume = "136",
    pages = "387--391",
    year = "1984"
}

@article{Nelson:1984hg,
    author = "Nelson, Ann E.",
    title = "{Calculation of $\theta$ Barr}",
    reportNumber = "HUTP-84/A022",
    doi = "10.1016/0370-2693(84)90827-X",
    journal = "Phys. Lett. B",
    volume = "143",
    pages = "165--170",
    year = "1984"
}

@article{Campbell:2012zzb,
    author = "Campbell, C. J. and Radnaev, A. G. and Kuzmich, A. and Dzuba, V. A. and Flambaum, V. V. and Derevianko, A.",
    title = "{A Single-Ion Nuclear Clock for Metrology at the 19th Decimal Place}",
    eprint = "1110.2490",
    archivePrefix = "arXiv",
    primaryClass = "physics.atom-ph",
    doi = "10.1103/PhysRevLett.108.120802",
    journal = "Phys. Rev. Lett.",
    volume = "108",
    pages = "120802",
    year = "2012"
}

@article{Peik:2020cwm,
    author = "Peik, E. and Schumm, T. and Safronova, M. S. and P\'alffy, A. and Weitenberg, J. and Thirolf, P. G.",
    title = "{Nuclear clocks for testing fundamental physics}",
    eprint = "2012.09304",
    archivePrefix = "arXiv",
    primaryClass = "quant-ph",
    doi = "10.1088/2058-9565/abe9c2",
    journal = "Quantum Sci. Technol.",
    volume = "6",
    number = "3",
    pages = "034002",
    year = "2021"
}

@article{Preskill:1982cy,
    author = "Preskill, John and Wise, Mark B. and Wilczek, Frank",
    editor = "Srednicki, M. A.",
    title = "{Cosmology of the Invisible Axion}",
    reportNumber = "HUTP-82-A048, NSF-ITP-82-103",
    doi = "10.1016/0370-2693(83)90637-8",
    journal = "Phys. Lett. B",
    volume = "120",
    pages = "127--132",
    year = "1983"
}

@article{MICROSCOPE:2022doy,
    author = "Touboul, Pierre and others",
    collaboration = "MICROSCOPE",
    title = "{MICROSCOPE Mission: Final Results of the Test of the Equivalence Principle}",
    eprint = "2209.15487",
    archivePrefix = "arXiv",
    primaryClass = "gr-qc",
    doi = "10.1103/PhysRevLett.129.121102",
    journal = "Phys. Rev. Lett.",
    volume = "129",
    number = "12",
    pages = "121102",
    year = "2022"
}

@article{Perez:2020dbw,
    author = "Perez, Gilad and Shalit, Aviv",
    title = "{High quality Nelson-Barr solution to the strong CP problem with $\theta=\pi$}",
    eprint = "2010.02891",
    archivePrefix = "arXiv",
    primaryClass = "hep-ph",
    doi = "10.1007/JHEP02(2021)118",
    journal = "JHEP",
    volume = "02",
    pages = "118",
    year = "2021"
}

@article{Vecchi:2014hpa,
    author = "Vecchi, Luca",
    title = "{Spontaneous CP violation and the strong CP problem}",
    eprint = "1412.3805",
    archivePrefix = "arXiv",
    primaryClass = "hep-ph",
    doi = "10.1007/JHEP04(2017)149",
    journal = "JHEP",
    volume = "04",
    pages = "149",
    year = "2017"
}

@article{Kamionkowski:1992mf,
    author = "Kamionkowski, Marc and March-Russell, John",
    title = "{Planck scale physics and the Peccei-Quinn mechanism}",
    eprint = "hep-th/9202003",
    archivePrefix = "arXiv",
    reportNumber = "IASSNS-HEP-92-9, PUPT-92-1309",
    doi = "10.1016/0370-2693(92)90492-M",
    journal = "Phys. Lett. B",
    volume = "282",
    pages = "137--141",
    year = "1992"
}

@article{Randall:1992ut,
    author = "Randall, Lisa",
    title = "{Composite axion models and Planck scale physics}",
    reportNumber = "MIT-CTP-2074",
    doi = "10.1016/0370-2693(92)91928-3",
    journal = "Phys. Lett. B",
    volume = "284",
    pages = "77--80",
    year = "1992"
}

@article{Holman:1992us,
    author = "Holman, Richard and Hsu, Stephen D. H. and Kephart, Thomas W. and Kolb, Edward W. and Watkins, Richard and Widrow, Lawrence M.",
    title = "{Solutions to the strong CP problem in a world with gravity}",
    eprint = "hep-ph/9203206",
    archivePrefix = "arXiv",
    reportNumber = "NSF-ITP-92-06, CMU-HEP92-05, FERMILAB-PUB-92-034-A, HUTP-92-A011, VAND-TH-92-2",
    doi = "10.1016/0370-2693(92)90491-L",
    journal = "Phys. Lett. B",
    volume = "282",
    pages = "132--136",
    year = "1992"
}

@article{Barr:1992qq,
    author = "Barr, Stephen M. and Seckel, D.",
    title = "{Planck scale corrections to axion models}",
    reportNumber = "BA-92-11",
    doi = "10.1103/PhysRevD.46.539",
    journal = "Phys. Rev. D",
    volume = "46",
    pages = "539--549",
    year = "1992"
}

@article{Asadi:2022vys,
    author = "Asadi, Pouya and Homiller, Samuel and Lu, Qianshu and Reece, Matthew",
    title = "{Chiral Nelson-Barr models: Quality and cosmology}",
    eprint = "2212.03882",
    archivePrefix = "arXiv",
    primaryClass = "hep-ph",
    doi = "10.1103/PhysRevD.107.115012",
    journal = "Phys. Rev. D",
    volume = "107",
    number = "11",
    pages = "115012",
    year = "2023"
}

@article{Banerjee:2022wzk,
    author = "Banerjee, Abhishek and Eby, Joshua and Perez, Gilad",
    title = "{From axion quality and naturalness problems to a high-quality ZN QCD relaxion}",
    eprint = "2210.05690",
    archivePrefix = "arXiv",
    primaryClass = "hep-ph",
    doi = "10.1103/PhysRevD.107.115011",
    journal = "Phys. Rev. D",
    volume = "107",
    number = "11",
    pages = "115011",
    year = "2023"
}

\end{document}